\documentclass[prb,letterpaper,twocolumn,noshowpacs,preprintnumbers,amsmath,amssymb,showkeys]{revtex4}

\usepackage{epsfig}
\usepackage{graphicx}
\usepackage{dcolumn}
\usepackage{bm}
\usepackage{bbm}

\begin{document}


\title{A Path Integral Approach for Time-Dependent Hamiltonians with Applications to
  Derivatives Pricing}

\author{Mark Stedman}  \author{Luca Capriotti}

\affiliation{%
Department of Mathematics and Department of Industrial Engineering and Operations Research, Columbia University, New York, New York 10027, United States of America \\
}%

\date{\today}

\begin{abstract}
We generalize a semi-classical path integral approach originally introduced 
by Giachetti and Tognetti [Phys. Rev. Lett. {\bf 55}, 912 (1985)] and Feynman and Kleinert [Phys. Rev. A {\bf 34}, 5080 (1986)] 
to time-dependent Hamiltonians, thus extending the scope of the method to the pricing of financial derivatives.
We illustrate the accuracy of the approach by presenting results for the well-known, but
analytically intractable, Black-Karasinski model for the dynamics of interest rates. 
The accuracy and  computational efficiency of this path integral approach makes it a viable
alternative to fully-numerical schemes for a variety of applications in derivatives pricing. 
\end{abstract}


\maketitle

\section{Introduction}

The formal connection between  Euclidean path integrals 
\cite{FeynmanRMP1948, feynman1998statistical} and the pricing of financial derivatives has been known
for some time~\citep{Dash1988, Dash1989, Linetsky1997, BennatiRosaClot1999}.  
In particular, in a seminal paper, Bennati {\em et al.} \cite{BennatiRosaClot1999} showed that
the solution of stochastic differential, i.e., Langevin,
equations can be expressed in terms of a path integral similar to the one describing the quantum
statistical mechanics of condensed matter systems.


Over the years, the path integral approach has been shown to be a useful analytical
\cite{MONTAGNA2002450, LWTF08, LLT11, BAAQUIE20121408} and numerical tool
\cite{CAPUOZZO2021126231} for option pricing which complements the traditional methods, stemming
from stochastic calculus \cite{karatzas1991brownian}, that state the pricing problem in terms
of partial differential equations (PDEs).  

The different view point adopted in the path integral framework allows for the possibility of borrowing from the extensive physics literature on the topic in order to develop accurate approximation schemes 
that are not otherwise available, or known, in traditional formulations of mathematical finance\cite{BennatiRosaClot1999, Kakushadze2015, Capriotti2006}.  

Among these, {\em semi-classical} approaches \cite{kleinert2009path}, corresponding in a Langevin setting to the limit 
of small intensity of the white noise, play a central role. These  approximations  can be developed in several ways\cite{Wentzel1926, Kramers1926, Brillouin1926,Wigner1932,Kirkwood1933,FujiwaraOW1982,HilleryCSW1984}  which, while sharing the same limiting behavior, lead to genuinely different results.   
Among semi-classical approximations, a prominent place is occupied by the so-called {\em
  effective potential} methods \cite{FeynmanHibbs,feynman1998statistical,kleinert2009path}
based, in the language of Wilson's renormalization group, on `integrating out' the fluctuations around a `classical' trajectory.  Although exact in principle, the calculation can be performed only at some level of approximation, using a perturbation scheme in which the choice of the unperturbed system plays a crucial role in the quality of the approximation. 

A particularly successful effective potential approximation is the one arising from a simple and
nice idea originally due to Feynman \cite{feynman1998statistical} and developed independently by Giachetti and Tognetti \cite{GiachettiTognetti1985} and Feynman and Kleinert \cite{FeynmanKleinert1986} (GTFK), which is based on a self-consistent (non-local) harmonic approximation of the effective potential in a sense that will become clear in the following sections.


The most appealing aspect of this approach is that the classical behavior is fully accounted for
by the GTFK potential.  This opened a way to tackle challenging quantum systems whose classical
analogues were known to be characterized by nonlinear excitations, e.g., those dubbed solitons
in one dimension or vortices in two dimensions. As a result, GTFK methods have been
successfully employed to describe quantitatively the properties of condensed matter systems
showing these excitations, such as Sine-Gordon chains \cite{GiachettiTognetti1985} and two dimensional (2D)
anisotropic spin systems \cite{1995CTVVpra, CGTTVV95}. Other quantum systems that have been
succesfully treated by (suitable generalizations of) the same method are frustrated
antiferromagnets, e.g., the easy-plane Heisenberg model on the triangular lattice~\cite{CCTVV1999}, the 2D $J_1$-$J_2$ model~\cite{CapriottiFRT2004}, and the 2D Josephson-junction array
\cite{1997CRTVpre,2000CFTVprb}.


In this paper, we consider the application of the GTFK method to a class of interest rate
models  where the rate of interest applied instantaneously at time $t$, known as the `short rate', is of the form $r(t)=r(X(t),t)$,  with $X(t)$ following the non-linear diffusion process specified by the stochastic differential equation (SDE), 
\begin{equation}\label{eq:SDE}
dX(t) = \mu(X(t), t)\,dt + \sigma(X(t), t)\,dW(t)~,
\end{equation}
for $t>0$, where $\mu(X(t), t)$ and $\sigma(X(t), t)$  are the drift and volatility functions, respectively, $X(0) = x_{0}$, and $W(t)$ is a standard Brownian motion \footnote{
The SDE is intended as a short-hand notation of the corresponding Ito's integral \cite{BennatiRosaClot1999}.}.
The same formalism can be also applied to describe the credit worthiness of obligors, e.g., in the context of default 
intensity models \citep{o2010modelling}.  

Short-rate models are of paramount importance in financial modeling, providing the foundation of many approaches 
used for the pricing of both interest rate and credit derivatives
\citep{andersen2010interest,o2010modelling}. In particular, the class of
{\em affine models} introduced by Duffie and Kan~\citep{duffie} play a prominent
role.  Celebrated examples of this class include the
Vasicek \cite{Vasicek1977},  Hull and White \cite{hw}, and Cox, Ingersoll and Ross\cite{cir}
models.  These models, for which $r(t) = X(t)$
and the drift and the square of the volatility functions in Eq.~(\ref{eq:SDE}) are either
constant or linear in $X(t)$, are analytically tractable
with closed-form expressions for fundamental financial quantities such as zero-coupon bonds and European option prices.
Unfortunately, the availability of closed-form solutions comes at the price of less than
realistic properties of the underlying rate dynamics. 

On the other hand, more realistic models are less analytically tractable than the affine models. As a result, although widely used in practice, 
their implementations rely on computationally-intensive  numerical methods, e.g., PDE  or Monte Carlo, for the calculation of  bond and derivatives prices. 
This is particularly onerous in the context of  multi-factor problems, notably those
involving the calculation of valuation adjustments (XVA)~\cite{gregory}, that are currently of
great interest in financial engineering. 

In this context, 
reliable analytical approximations are particularly important to reduce the numerical burden associated with these computations.
%
This has led several 
authors to propose approximation schemes that are based on various expansion
techniques \cite{AntonovSpector2011, daniluk2016, EEBK, horvath2018analytic}.  However, the
applicability of these expansions tends to be limited by their finite convergence
radius~\cite{capriotti2020path}.

Recently, the original formulation of the GTFK approximation was used to develop an accurate and easy-to-compute semi-analytical 
approximation for derivatives pricing problems where the underlying asset follows the form (\ref{eq:SDE}) with {\em time independent} drift and volatility \cite{capriotti2020path}.
However, the original GTFK approximation is restricted to Hamiltonians with no explicit time
dependence and, as such, it cannot be applied as an approximation method for SDEs with explicit time dependence 
in their coefficients.
This limits its usefulness for the pricing of interest rate and credit derivatives
since the time-dependence of the drift $\mu(x, t)$ and volatility $\sigma(x, t)$ in Eq.~(\ref{eq:SDE}), is a key model feature that is necessary to
calibrate the dynamics to the initial term structure of interest or hazard rates and traded options prices.

In this paper, we generalize the GTFK approximation to {\em time-dependent} Hamiltonians, thus extending the scope of the method.  
In addition to being exact for Gaussian models, we find that the new approximation provides
accurate results for the widely-used Black-Karasinski (BK) model \cite{bk}, for which the
zero-coupon bond prices are not known in closed form~\cite{andersen2010interest},
even in regimes of high volatility and multi-year time horizons.  The accuracy and computational
efficiency of the new approximation compares favourably with previously-presented approximation
schemes and  makes it a viable alternative to fully-numerical schemes for a variety of applications
in derivatives pricing.

The remainder of this paper is organized as follows. We begin by introducing the formalism of Arrow-Debreu prices and making the connection with Euclidean path integrals in Section \ref{sec::introduction}.  We show that, in general, the path integral that one needs to solve involves a time-dependent Hamiltonian which makes traditional effective potential 
methods inapplicable.  
As a first step in generalizing the GTFK approach to this class of Hamiltonian, we derive the density matrix for the forced harmonic oscillator with time-dependent
parameters in Section~\ref{sec:FHODensityMatrix}.  In Section~\ref{sec:GeneralizedGTFK}, we then use this result to derive the 
generalization of the GTFK approximation to time-dependent Hamiltonians.  We demonstrate the effectiveness of the new approximation in the case of
generalized short rate models in Section~\ref{sec:GeneralizedShortRate}. Section~\ref{sec:Conclusion} concludes.

\section{Arrow-Debreu densities and Path Integrals}
\label{sec::introduction}

In this paper we will focus on developing approximations of the so-called (generalized) {\em
  Arrow-Debreu (AD) densities} \cite{andersen2010interest, karatzas1991brownian}, also known as
Green's functions, which are the fundamental building  blocks of derivatives pricing. For a random variable following Eq. (1), these are defined as 
%
%
\begin{align} 
  \psi_\lambda(x_{T}, T&;x_{t}, t) = \nonumber \\
  &\mathbb{E}_{t}\Big[\delta(X(T) - x_{T}) e^{-\lambda \int_{t}^{T} r(X(s), s)\,ds}\Big]~,
 \label{eq:ADdens}
\end{align}
where $\mathbb{E}_{t}[\cdot] = \mathbb{E}[\cdot\,|\,X(t) = x_{t}]$, $\lambda$ is a real number, $\delta(\cdot)$ is the standard Dirac $\delta$ function,
and $r(x, t)$ is the short rate of interest for the state $(x, t)$.
This quantity, for $\lambda = 0$, gives the transition density, specifying the probability for the random variable following
Eq.~(\ref{eq:SDE}) to assume a value in an infinitesimal interval around $x_{T}$ at time $T$, given that it was at $x_{t}$ at time $t$, such that
\begin{equation}
  \int_A \psi_{0}(x_{T}, T; x_{t}, t)\,dx_{T} \equiv \mathbb{P}\left[X(T) \in A \,|\, X(t) = x_{t}
    \right]~.
\end{equation}
The value at time $t$ of a European option with expiry $T \ge t$ and an arbitrary payout
of the form $P(x)$,
\begin{equation}
V(t) = \mathbb{E}_{t}\Big[e^{-\int_{t}^{T} r(X(s), s)\,ds} P(X(T)) \Big],
\end{equation}
can be obtained by integrating  the product of the payout function and the ($\lambda = 1$) AD
density over all the possible values of $x_{T}$, i.e.,
\begin{equation}
V(t) = \int \psi_1(x_{T}, T; x_{t}, t) P(x_T)\,dx_{T}~.
\label{eq:PsiPricing}
\end{equation}
In particular, setting $P\equiv 1$
the moment generating function for 
the random process $\int_{t}^T r(X(s),s)\,ds$  
can be obtained 
\begin{equation}\label{eq:zeroad}
Z_\lambda(t, T; r_{t}) = \int \psi_{\lambda}(x_{T}, T; x_{t}, t)\,dx_{T}~,
\end{equation}
which, for $\lambda=1$, gives the time-$t$ value of a zero-coupon bond with maturity $T$, $Z( t, T ; r_{t})$ \cite{andersen2010interest}.


Within the path integral formulation, applying the prescription in Bennati {\em et al.}\cite{BennatiRosaClot1999}
to the SDE (\ref{eq:SDE}), one can show that the solution of the 
Fokker-Planck equation satisfied by Arrow-Debreu densities (\ref{eq:ADdens}) can be written as
\begin{equation}
  \psi_{\lambda}(x_{T}, T; x_{t}, t) = \int_{x(t) = x_{t}}^{x(T) = x_{T}} \hspace{-0.1cm} 
  e^{-S[x(u)]} \,\, {\cal D}[x(u)]~,
\label{eq:PsiPathInt}
\end{equation}
where the path integral, $\int_{x(t) =x_t}^{x(T) = x_T} \hspace{-0.1cm}{\cal D} [x(u)] \ldots$,   is defined over all paths $x(u)$ such that $x(t) = x_t$ and $x(T) = x_T$\cite{BennatiRosaClot1999}. Here 
\begin{equation}
  S[x(u)] = \int_{t}^{T} H(x(u), \dot{x}(u), u)\,du
\end{equation}
with
\begin{eqnarray}
  H(x, \dot{x}, u) & = & \frac{1}{2\sigma^{2}(u, x)}\left[
    \dot{x} - \mu(x, u)\right]^{2} + \frac{1}{2}\partial_{x}\mu(x, u) \nonumber\\
    & + & \lambda r(x, u)~.
\label{eq:PsiGenHamiltonian}
\end{eqnarray}

The key observation \cite{BennatiRosaClot1999} is that Eq.~(\ref{eq:PsiPathInt}) has the same
form as the Euclidean path integral 
representing the  
density matrix in the coordinate representation, namely
\begin{equation}
  \rho(b, a) = \int_{x(u_{a})=x_{a}}^{x(u_{b})=x_{b}} \hspace{-0.1cm} 
  e^{-S[x(u)]/\hbar }\, \,{\cal D} [x(u)]~,
\label{eq:DensityMatrix}
\end{equation}
where $a = (u_{a}, x_{a})$, $b = (u_{b}, x_{b})$, 
\begin{equation}
  S[x(u)] = \int_{u_{a}}^{u_{b}} H(x(u), \dot{x}(u), u)\,du~,
\label{eq:Action}
\end{equation}
$H(x(u), \dot{x}(u), u)$ is the Hamiltonian, and $\hbar$ is the reduced Planck's constant.

Given this formal equivalence, we can hope to be able to use known approximation methods in quantum mechanics to develop accurate 
approximations of Arrow-Debreu densities. The main hurdle to overcome is the time dependency in the Hamiltonian (\ref{eq:PsiGenHamiltonian}) 
which typically is not encountered in physics. 

In the following we will generalize the GTFK effective potential method \cite{GiachettiTognetti1985,FeynmanKleinert1986} to time-dependent Hamiltonians. We begin by deriving
the density matrix for the forced harmonic oscillator with time-dependent parameters, upon which we will build our generalization of the GTFK approach.


\section{Density matrix for the forced harmonic oscillator with time-dependent parameters}
\label{sec:FHODensityMatrix}

The Hamiltonian for the forced harmonic oscillator with time-dependent parameters is
\begin{equation}
  H(x, \dot{x}, u) = \frac{m(u)}{2}\dot{x}^{2} + \frac{m(u)}{2}\omega^{2}(u)x^{2}
  + \gamma(u)x + w(u)~,
\label{eq:FHOHamiltonian}
\end{equation}
where $m(u)$ is the mass, $\omega(u)$ is the angular frequency, $\gamma(u)$ is an
external force, and $w(u)$ is a shift.

As shown in  Appendix \ref{sec:tdfho}, since Eq.~(\ref{eq:FHOHamiltonian}) is of the quadratic form
(\ref{eq:GenHamiltonian}), we may apply the results of Appendix \ref{sec:GaussianPathInt} to
obtain the density matrix in closed form.

Specifically, the density matrix can be expressed in terms of a function $\dot \nu(u)$ as  
\begin{equation}
  \rho(b,  a) = \left(\frac{\mu_{b}\mu_{a}\sqrt{\dot{\nu}_{b}\dot{\nu}_{a}}}
      {2\pi \hbar \sinh\left(\nu_{b} - \nu_{a}\right)}\right)^{1/2}
    e^{-S_{\mathrm{cl}}(b, a) / \hbar}~,
\label{eq:FHODensityMatrix}
\end{equation}
where $\mu(u) \equiv \sqrt{m(u)}$, $\mu_{b} \equiv \mu(u_{b})$, $\nu_{b} \equiv \nu(u_{b})$,
etc.,
\begin{equation}
    \nu(u^{\prime\prime}) - \nu(u^{\prime}) = \int_{u^{\prime}}^{u^{\prime\prime}} \dot{\nu}(u)\,du~,
\end{equation}
and  the classical action is
\begin{eqnarray}
  S_{\mathrm{cl}}(b, a) & = & -\frac{1}{2}\left[
    \left(\frac{\ddot{\nu}_{b}}{2\dot{\nu}^{2}_{b}}
      + \frac{\dot{\mu}_{b}}{\mu_{b}\dot{\nu}_{b}}\right) \tilde{x}_{b}^{2}
      \right. \nonumber\\
     & - & \left.\left(\frac{\ddot{\nu}_{a}}{2\dot{\nu}^{2}_{a}}
      + \frac{\dot{\mu}_{a}}{\mu_{a}\dot{\nu}_{a}}\right) \tilde{x}_{a}^{2}
      \right] \nonumber\\
  & + & \frac{1}{2\sinh\left(\nu_{b} - \nu_{a}\right)}
  \left[\left(\tilde{x}_{b}^{2} + \tilde{x}_{a}^{2}\right)
    \cosh\left(\nu_{b} - \nu_{a}\right) \right.\nonumber\\
    & - & 2\tilde{x}_{b}\tilde{x}_{a}
    + \left. 2\tilde{x}_{b} \tilde{\Gamma}_{a}
    + 2 \tilde{x}_{a} \tilde{\Gamma}_{b}
    - 2 \tilde{\Gamma}_{ab}
    \right] \nonumber\\
    & + & \int_{u_{a}}^{u_{b}} w(u)\,du~,
\label{eq:FHOS_clb}
\end{eqnarray}
where
\begin{eqnarray}
  \tilde{x}_{b} & = & \mu_{b}\sqrt{\dot{\nu}_{b}} x_{b} \nonumber\\
  \tilde{x}_{a} & = & \mu_{a}\sqrt{\dot{\nu}_{a}} x_{a}~,
\end{eqnarray}
and
\begin{eqnarray}
  \tilde{\Gamma}_{a} & \equiv &
  \int_{u_{a}}^{u_{b}}\tilde{\gamma}(u)\sinh\left(\nu(u) - \nu_{a}\right)\,du \nonumber\\
  \tilde{\Gamma}_{b} & \equiv &
    \int_{u_{a}}^{u_{b}}\tilde{\gamma}(u)\sinh\left(\nu_{b} - \nu(u)\right)\,du \nonumber\\
  \tilde{\Gamma}_{ab} & \equiv &
    \int_{u_{a}}^{u_{b}}\tilde{\gamma}(u)\sinh\left(\nu_{b} - \nu(u)\right) \nonumber\\
    & \times & \int_{u_{a}}^{u}\tilde{\gamma}(u^{\prime})
    \sinh\left(\nu(u^{\prime}) - \nu_{a}\right)\,du^{\prime}\,du~,
\label{eq:GammaIntDef}
\end{eqnarray}
with
\begin{equation}
  \tilde{\gamma}(u)  =  \frac{\gamma(u)}{\mu(u)\sqrt{\dot{\nu}(u)}}~.
\end{equation}
Finally, the function $\dot \nu(u)$ can be obtained setting 
\begin{equation}
  h(u) = \dot{\nu}^{-1/2}(u)
\label{eq:h}
\end{equation}
and solving a Pinney equation \cite{Pinney1950} of the form
\begin{equation}
  \ddot{h}(u) - \left(\omega^{2}(u) + \frac{\ddot{\mu}(u)}{\mu(u)}\right)h(u)
  + \frac{1}{h^{3}(u)} = 0~.
\label{eq:Pinney}
\end{equation}

Note that in the case that both $\mu(u)$ and $\omega(u)$ are constant, Eq.~(\ref{eq:Pinney})
admits the simple solution $h(u) = \dot{\nu}^{-1/2}(u) = \omega^{-1/2}$ and the expression in 
Eq.~(\ref{eq:FHODensityMatrix}) reduces to the well-known Green's function for the forced harmonic
oscilator with constant mass and frequency under the substitution
$t = -i u$, see, e.g., Ref.~\onlinecite{FeynmanHibbs}.

\section{A generalized GTFK approximation}
\label{sec:GeneralizedGTFK}

The GTFK method \cite{GiachettiTognetti1985,FeynmanKleinert1986} is based on a self-consistent, non-local, harmonic approximation of the
effective potential.  Specifically, paths in the functional integration in
(\ref{eq:DensityMatrix}) are classified according to  their average point, defined as the
functional
\begin{equation}
  \bar x[x(u)] = \frac{1}{u_{b} - u_{a}} \int_{u_{a}}^{u_{b}} x(u)\,du~,
\end{equation}
so that each equivalence class is labelled by a real number $\bar{x}$ representing the common
average point.  Hence, we can factor out an ordinary integral over $\bar{x}$ in
Eq.~(\ref{eq:DensityMatrix}) to obtain
\begin{equation}
  \rho(b, a) =  \int \bar{\rho}(b, a; \bar{x})\,d\bar{x}~, 
\label{eq:Rho}
\end{equation}
in which the {\em reduced density matrix}
\begin{eqnarray}
  \bar{\rho}(b, a; \bar{x}) & = & \int_{x(a) = x_{a}}^{x(b) = x_{b}} \hspace{-0.1cm}
  \delta\left(\bar{x} - \frac{1}{u_{b} - u_{a}} \int_{u_{a}}^{u_{b}} x(u)\,du \right) \nonumber\\
  & \times & e^{-S[x(u)] / \hbar}\, \,{\cal D} [x(u)]~,
\label{eq:ReducedDensityMatrix}
\end{eqnarray}
represents the contribution to the path integral in Eq.~(\ref{eq:DensityMatrix}) due to the paths
that have average point $\bar{x}$. 

As the path integral in Eq.~(\ref{eq:ReducedDensityMatrix}) is limited to paths
that belong to the same class, we may develop a specialized approximation for each class.  In
particular, we can approximate the action (\ref{eq:Action}) with a quadratic action in the
displacement from the average point $\bar{x}$.

The original GTFK method\cite{GiachettiTognetti1985} is based on a trial Hamiltonian of the form
\begin{equation}
H(x, \dot x ,u ;\bar x) =  \frac{m}{2} \dot x^{2} + V(x; \bar{x}) 
\end{equation}
where
\begin{equation}\label{eq.trialpot}
V(x; \bar{x}) = \frac{m \omega^2(\bar x)}{2} (x-\bar x)^2 + w(\bar x)~,
\end{equation}

In this section, we generalize the GTFK approximation to use a {\em time-dependent}
quadratic trial Hamiltonian of the form
\begin{equation}
  H_{0}(x, \dot{x}, u; \bar{x}) = \frac{m(u)}{2}\dot{x}^{2} + V_{0}(x, u; \bar{x})~.
\end{equation}
where the trial potential is
\begin{eqnarray}
  V_{0}(x, u; \bar{x}) & = &
  \frac{m(u)}{2}\omega^{2}(u; \bar{x})\left(x - \bar{x}\right)^{2}
  +  \gamma(u; \bar{x})\left(x - \bar{x}\right)
  \nonumber\\
  & + & w(u; \bar{x})~,
\label{eq:V_0}
\end{eqnarray}
and $\bar{x}$ is a constant parameter, thus extending the potential scope of application of
the method to problems with a time-dependent Hamiltonian.

Let us, therefore, consider the action
\begin{equation}
  S_{0}[x(u)] = \int_{u_{a}}^{u_{b}} H_{0}(x(u), \dot{x}(u), u; \bar{x})\,du~.
\end{equation}
Using the Fourier representation of the Dirac $\delta$ function, we can write the reduced
density matrix as
\begin{eqnarray}
  \bar{\rho}_{0}(b, a; \bar{x}) & = & \frac{u_{b} - u_{a}}{2\pi \hbar} \int_{-\infty}^{\infty}
  \int_{x(a) = x_{a}}^{x(b) = x_{b}} \hspace{-0.1cm} e^{-S_{0}[x(u)] / \hbar}
  \nonumber\\
  & \times & \exp \left[
    -\frac{iy}{\hbar}\int_{u_{a}}^{u_{b}} \left(x(u) - \bar{x}\right)\,du \right]
  \nonumber\\
  & \times & {\cal D} [x(u)]\,dy \\
  & = & \frac{u_{b} - u_{a}}{2\pi \hbar} \int_{-\infty}^{\infty}
  \int_{x(a) = x_{a}}^{x(b) = x_{b}} \hspace{-0.1cm}
  e^{-S_{1}[x(u)] / \hbar}
  \nonumber\\
  & \times & {\cal D} [x(u)]\,dy~,
\end{eqnarray}
where
\begin{equation}
  S_{1}[x(u)] = \int_{u_{a}}^{u_{b}} H_{1}(x(u), \dot{x}(u), u; \bar{x}, y)\,du
\end{equation}
and
\begin{eqnarray}
  H_{1}(x, \dot{x}, u; \bar{x}, y) & = & \frac{m(u)}{2}\dot{x}^{2}
  + \frac{m(u)}{2}\omega^{2}(u; \bar{x})\left(x - \bar{x}\right)^{2}
  \nonumber\\
  & + & \left[\gamma(u, \bar{x}) + iy\right]\left(x - \bar{x}\right) + w(u; \bar{x})~.
  \nonumber\\
\end{eqnarray}
Changing variables to $x^{\prime} = x - \bar{x}$ and noting that $\dot{x}^{\prime} = \dot{x}$
since $\bar{x}$ is constant, we see that, in terms of $x^{\prime}$, $H_{1}$ is the Hamiltonian
for the forced harmonic oscillator with time-dependent parameters.  Hence, we may use the
anayltic continuation of the results of Section~\ref{sec:FHODensityMatrix} to write
\begin{eqnarray}
  \bar{\rho}_{0}(b, a; \bar{x}) & = & \frac{u_{b} - u_{a}}{2\pi \hbar}
  \left(\frac{\mu_{b}\mu_{a}\sqrt{\dot{\nu}_{b}\dot{\nu}_{a}}}
       {2\pi \hbar \sinh\left(\nu_{b} - \nu_{a}\right)}\right)^{1/2}
       \nonumber\\
       & \times & \int_{-\infty}^{\infty} e^{-S_{\mathrm{cl}}(b^{\prime}, a^{\prime}; y) / \hbar}\,dy~,  
\label{eq:ReducedDensityMatrix2}
\end{eqnarray}
where $x^{\prime}_{b} = x_{b} - \bar{x}$, $x^{\prime}_{a} = x_{a} - \bar{x}$,
\begin{eqnarray}
  \tilde{x}^{\prime}_{b} & = & \mu_{b}\sqrt{\dot{\nu}_{b}} x^{\prime}_{b} \nonumber\\
  \tilde{x}^{\prime}_{a} & = & \mu_{a}\sqrt{\dot{\nu}_{a}} x^{\prime}_{a}~,
\end{eqnarray}
and, suppressing the dependence on $\bar{x}$ to lighten notation,
\begin{eqnarray}
  S_{\mathrm{cl}}(b^{\prime}, a^{\prime}; y) & = & -\frac{1}{2}\left[
    \left(\frac{\ddot{\nu}_{b}}{2\dot{\nu}^{2}_{b}}
      + \frac{\dot{\mu}_{b}}{\mu_{b}\dot{\nu}_{b}}\right) \tilde{x}_{b}^{\prime2}
      \right. \nonumber\\
    & - &  \left.\left(\frac{\ddot{\nu}_{a}}{2\dot{\nu}^{2}_{a}}
      + \frac{\dot{\mu}_{a}}{\mu_{a}\dot{\nu}_{a}}\right) \tilde{x}_{a}^{\prime2}
      \right] \nonumber\\
  & + & \frac{1}{2\sinh\left(\nu_{b} - \nu_{a}\right)}
  \nonumber\\
  & \times & \Big[\left(\tilde{x}_{b}^{\prime2} + \tilde{x}_{a}^{\prime2}\right)
    \cosh\left(\nu_{b} - \nu_{a}\right)
    - 2\tilde{x}^{\prime}_{b}\tilde{x}^{\prime}_{a} \nonumber\\
    & + & 2\tilde{x}^{\prime}_{b} \left(\tilde{\Gamma}_{a} + iy\Omega_{a}\right)
    + 2 \tilde{x}^{\prime}_{a} \left(\tilde{\Gamma}_{b} + iy\Omega_{b}\right)
    \nonumber\\
    & - & \left. 2 \left(
    \tilde{\Gamma}_{ab} + iy\left(I_{1} + I_{2}\right) - y^{2}\Omega_{ab}\right)
    \right] \nonumber\\
    & + & \int_{u_{a}}^{u_{b}} w(u)\,du
\end{eqnarray}
with the definitions
\begin{eqnarray}
  \Omega_{a} & \equiv &
  \int_{u_{a}}^{u_{b}} \mu^{-1}(u)\nu^{-1/2}(u)
  \sinh\left(\nu(u) - \nu_{a}\right)\,du \nonumber\\
  \Omega_{b} & \equiv &
  \int_{u_{a}}^{u_{b}} \mu^{-1}(u)\nu^{-1/2}(u)
  \sinh\left(\nu_{b} - \nu(u)\right)\,du \nonumber\\
  \Omega_{ab} & \equiv &
  \int_{u_{a}}^{u_{b}} \mu^{-1}(u)\nu^{-1/2}(u) \sinh\left(\nu_{b} - \nu(u)\right)
  \nonumber\\
  & \times & \int_{u_{a}}^{u} \mu^{-1}(u)\nu^{-1/2}(u)
  \sinh\left(\nu(u^{\prime}) - \nu_{a}\right)\,du^{\prime}\,du \nonumber \\
  I_{1} & \equiv &
  \int_{u_{a}}^{u_{b}} \tilde{\gamma}(u) \sinh\left(\nu_{b} - \nu(u)\right)
  \nonumber\\
  & \times & \int_{u_{a}}^{u} \mu^{-1}(u)\nu^{-1/2}(u)
  \sinh\left(\nu(u^{\prime}) - \nu_{a}\right)\,du^{\prime}\,du \nonumber \\
  I_{2} & \equiv &
  \int_{u_{a}}^{u_{b}} \mu^{-1}(u)\nu^{-1/2}(u) \sinh\left(\nu_{b} - \nu(u)\right)
  \nonumber\\
  & \times & \int_{u_{a}}^{u} \tilde{\gamma}(u^{\prime})
  \sinh\left(\nu(u^{\prime}) - \nu_{a}\right)\,du^{\prime}\,du~.
\label{eq:OmegaIIntDef}
\end{eqnarray}

We see that (\ref{eq:ReducedDensityMatrix2}) is a Gaussian integral in $y$ of the form
(\ref{eq:StandardGaussian}).  Performing the integration and simplifying, we obtain
\begin{eqnarray}
  \bar{\rho}_{0}(b, a; \bar{x}) & = & \frac{u_{b} - u_{a}}{2\pi \hbar}
  \left(\frac{\mu_{b}\mu_{a}\sqrt{\dot{\nu}_{b}\dot{\nu}_{a}}}
       {2 \Omega_{ab}}\right)^{1/2}
       \nonumber\\
       & \times & \exp\left\{\frac{1}{2\hbar}\left[
    \left(\frac{\ddot{\nu}_{b}}{2\dot{\nu}^{2}_{b}}
      + \frac{\dot{\mu}_{b}}{\mu_{b}\dot{\nu}_{b}}\right) \tilde{x}_{b}^{\prime2}
      \right.\right.\nonumber\\
    & - &  \left.\left(\frac{\ddot{\nu}_{a}}{2\dot{\nu}^{2}_{a}}
      + \frac{\dot{\mu}_{a}}{\mu_{a}\dot{\nu}_{a}}\right) \tilde{x}_{a}^{\prime2}
      \right] \nonumber\\
  & - & \frac{1}{2\hbar\sinh\left(\nu_{b} - \nu_{a}\right)}
  \Big[\left(\tilde{x}_{b}^{\prime2} + \tilde{x}_{a}^{\prime2}\right)
    \cosh\left(\nu_{b} - \nu_{a}\right)
    \nonumber\\
    & - & 2\tilde{x}^{\prime}_{b}\tilde{x}^{\prime}_{a}
    + 2\tilde{x}^{\prime}_{b} \tilde{\Gamma}_{a}
    + 2 \tilde{x}^{\prime}_{a} \tilde{\Gamma}_{b} - 2 \tilde{\Gamma}_{ab}
    \nonumber\\
    & + & \left. \frac{1}{2 \Omega_{ab}} \left(
    \tilde{x}^{\prime}_{b} \Omega_{a} + \tilde{x}^{\prime}_{a} \Omega_{b}
    -\left(I_{1} + I_{2}\right)\right)^{2}
    \right] \nonumber\\
    & - & \left. \frac{1}{\hbar} \int_{u_{a}}^{u_{b}} w(u)\,du\right\}~.
\label{eq:ReducedDensityMatrix3}
\end{eqnarray}

Now consider the diagonal element $x_{b} = x_{a} = x$, 
$\bar{\rho}_{0}(x, x; \bar{x})$.  Collecting terms
in $x^{\prime}$ and completing the square we obtain the Gaussian form
\begin{eqnarray}
 \bar{\rho}_{0}(x, x; \bar{x}) & = & \frac{u_{b} - u_{a}}{\hbar}
 \left(\frac{\alpha\mu_{b}\mu_{a}\sqrt{\dot{\nu}_{b}\dot{\nu}_{a}}}
      {4\pi \Omega_{ab}}\right)^{1/2}
      \nonumber\\
      & \times & \exp\left\{
\frac{1}{\hbar\sinh\left(\nu_{b} - \nu_{a}\right)}
      \left[\tilde{\Gamma}_{ab}
        - \frac{\left(I_{1} + I_{2}\right)^{2}}{4 \Omega_{ab}}\right]
      \right. \nonumber \\
      & + & \left. \frac{\delta_{\gamma}^{2}}{2 \alpha}
      - \frac{1}{\hbar} \int_{u_{a}}^{u_{b}} w(u)\,du \right\} \nonumber \\
      & \times & \frac{1}{\sqrt{2 \pi \alpha}}
        \exp\left\{-\frac{1}{2 \alpha}\left(x^{\prime} + \delta_{\gamma}\right)^{2}\right\}~,
\label{eq:DiagonalReducedDensityMatrix}\end{eqnarray}
where we have defined
\begin{eqnarray}
  \alpha & \equiv & \frac{1}{2C} \nonumber \\
  \delta_{\gamma} & \equiv & \frac{D}{2C}
\label{eq:AlphaDeltaGamma}
\end{eqnarray}
with
\begin{eqnarray}
  C & = & -\frac{1}{2\hbar}\left[
    \left(\frac{\ddot{\nu}_{b}}{2\dot{\nu}_{b}}
      + \frac{\dot{\mu}_{b}}{\mu_{b}}\right)\mu^{2}_{b}
      - \left(\frac{\ddot{\nu}_{a}}{2\dot{\nu}_{a}}
      + \frac{\dot{\mu}_{a}}{\mu_{a}}\right)\mu^{2}_{a}
      \right] \nonumber\\
  & + & \frac{1}{2\hbar\sinh\left(\nu_{b} - \nu_{a}\right)}
  \Big[\left(\mu^{2}_{b}\dot{\nu}_{b} + \mu^{2}_{a}\dot{\nu}_{a}\right)
    \cosh\left(\nu_{b} - \nu_{a}\right) \nonumber \\
    & - & 2\mu_{b}\mu_{a}\sqrt{\dot{\nu}_{b}\dot{\nu}_{a}}
    + \left. \frac{1}{2 \Omega_{ab}} \left(
    \mu_{b}\sqrt{\dot{\nu}_{b}} \Omega_{a}
    + \mu_{a}\sqrt{\dot{\nu}_{a}} \Omega_{b}\right)^{2}\right]
    \nonumber\\
  D & = & \frac{1}{\hbar\sinh\left(\nu_{b} - \nu_{a}\right)}\left[
    \mu_{b}\sqrt{\dot{\nu}_{b}} \tilde{\Gamma}_{a}
    + \mu_{a}\sqrt{\dot{\nu}_{a}} \tilde{\Gamma}_{b} \right. \nonumber\\
    & - & \left. \frac{\left(I_{1} + I_{2}\right)}{2 \Omega_{ab}}\left(
    \mu_{b}\sqrt{\dot{\nu}_{b}} \Omega_{a}
    + \mu_{a}\sqrt{\dot{\nu}_{a}} \Omega_{b}\right) \right]~.
\label{eq:CD}
\end{eqnarray}

Given the reduced density matrix $\bar{\rho}_{0}(x, x; \bar{x})$, the expectation of an
observable ${\cal O}(\hat{x})$ may be written as
\begin{equation}
  \langle{\cal O}(\hat{x})\rangle = \frac{1}{{\cal Z}}
  \int \int {\cal O}(x) \bar{\rho}_{0}(x, x; \bar{x})\,dx\,d\bar{x}~,
\end{equation}
where ${\cal Z}$ is a normalization factor.  Writing
\begin{equation}
  \bar{\rho}_{0}(x, x; \bar{x}) = \rho(\bar{x}) \frac{1}{\sqrt{2 \pi \alpha}}
        \exp\left\{-\frac{1}{2 \alpha}\left(x^{\prime} + \delta_{\gamma}\right)^{2}\right\}~,
\end{equation}
where $\rho(\bar{x})$ is the leading factor in Eq.~(\ref{eq:DiagonalReducedDensityMatrix}), and
changing the inner integration variable to $\xi = x - \bar{x} + \delta_{\gamma}$ we have
\begin{eqnarray}
  \langle{\cal O}(\hat{x})\rangle & = & \frac{1}{{\cal Z}}
  \int \frac{\rho(\bar{x})}{\sqrt{2 \pi \alpha}}
    \int {\cal O}(\bar{x} - \delta_{\gamma} + \xi)e^{-\xi^{2} / 2\alpha}
  \,d\xi\,d\bar{x} \nonumber\\
  & = & \frac{1}{{\cal Z}}
  \int \rho(\bar{x}) \langle\langle {\cal O}(\bar{x} - \delta_{\gamma} + \xi) \rangle\rangle
  \,d\bar{x}~,
\end{eqnarray}
where we have defined
\begin{eqnarray}
  \langle\langle {\cal O}(\bar{x} - \delta_{\gamma} + \xi) \rangle\rangle & \equiv &
  \frac{1}{\sqrt{2 \pi \alpha}}
  \int {\cal O}(\bar{x} - \delta_{\gamma} + \xi)   e^{-\xi^{2} / 2\alpha} \,d\xi
  \nonumber \\
  & = & \left.e^{\frac{\alpha}{2}\partial^{2}_{x}}{\cal O}(x)\right|_{x = \bar{x} - \delta_{\gamma}}~.
\label{eq:GaussianExpectation}
\end{eqnarray}
Note that the final equality in Eq.~(\ref{eq:GaussianExpectation}) comes from expanding the
integrand around $\bar{x} - \delta_{\gamma}$ and substituting for the central moments of the
Gaussian distribution.

In particular, for the trial potential given by Eq.~(\ref{eq:V_0}), the expectation of $V_{0}$ and
its first two derivatives with respect to $x$ are 
\begin{eqnarray}
  \langle\langle V_{0}(\bar{x} - \delta_{\gamma} + \xi) \rangle\rangle & = &
  \frac{m(u)}{2}\omega^{2}(u; \bar{x})\left(\delta_{\gamma}^{2} + \alpha\right) \nonumber\\
  & - & \gamma(u; \bar{x})\delta_{\gamma} + w(u; \bar{x}) \nonumber\\
  \langle\langle V_{0}^{\prime}(\bar{x} - \delta_{\gamma} + \xi) \rangle\rangle & = &
  -m(u)\omega^{2}(u; \bar{x})\delta_{\gamma} +  \gamma(u; \bar{x}) \nonumber\\
  \langle\langle V_{0}^{\prime\prime}(\bar{x} - \delta_{\gamma} + \xi) \rangle\rangle & = &
  m(u)\omega^{2}(u; \bar{x})~.
\label{eq:SelfConsistentV_0}
\end{eqnarray}

Hence, given any potential $V(x, u)$, we can determine  the parameters
$\omega(u; \bar{x})$, $\gamma(u; \bar{x})$, and $w(u' \bar{x})$ of the trial potential $V_{0}$
that best approximate $V$ by setting
\begin{eqnarray}
  \langle\langle V(\bar{x} - \delta_{\gamma} + \xi) \rangle\rangle & = &
  \langle\langle V_{0}(\bar{x} - \delta_{\gamma} + \xi) \rangle\rangle \nonumber\\
  \langle\langle V^{\prime}(\bar{x} - \delta_{\gamma} + \xi) \rangle\rangle & = &
  \langle\langle V_{0}^{\prime}(\bar{x} - \delta_{\gamma} + \xi) \rangle\rangle \nonumber\\
  \langle\langle V^{\prime\prime}(\bar{x} - \delta_{\gamma} + \xi) \rangle\rangle & = &
  \langle\langle V_{0}^{\prime\prime}(\bar{x} - \delta_{\gamma} + \xi) \rangle\rangle~.
\label{eq:SelfConsistent}
\end{eqnarray}
This ensures that the expectation values of $V_{0}$ and of its second order expansion
according to the Gaussian probability distribution in Eq.~(\ref{eq:GaussianExpectation})  are
in agreement with those of $V$, for {\em every} value of $\bar x$. 

\subsection{Quadratic Hamiltonians}

Given the form of the chosen trial potential, we expect our GTFK approximation to be exact
for quadratic Hamiltonians.  Showing that this is indeed the case is an important check
on the consistency of the approximation.

As is apparent from (\ref{eq:GenS_cl}) and (\ref{eq:GenHomogeneousEqnOfMotion}), and may be
shown directly by integrating by parts the terms in $\dot{x}x$ and $\dot{x}$ that arise when
(\ref{eq:GenHamiltonian}) is inserted into (\ref{eq:Action}), the density matrix for any
quadratic Hamiltonian can be written formally as the density matrix for the forced harmonic
oscillator~(\ref{eq:FHODensityMatrix}) times a factor that depends only on the end points
provided that we redefine, $m(u)$, $\omega^{2}(u)$, and
$\gamma(u)$ such that
\begin{eqnarray}
  m(u) & = & 2a(u) \nonumber\\
  m(u)\omega^{2}(u) & = & 2\left(c(u) - \dot{b}(u)\right) \nonumber\\
  \gamma(u) & = & 2\left(e(u) - \dot{d}(u)\right)~.
\end{eqnarray}
Hence, it suffices to show that our GTFK approximation is exact for the Hamiltonian in Eq.~(\ref{eq:FHOHamiltonian}).

Applying the GTFK approximation to Eq.~(\ref{eq:FHOHamiltonian}) we have
\begin{eqnarray}
  \langle\langle V(\bar{x} - \delta_{\gamma} + \xi) \rangle\rangle & = &
  \frac{m(u)}{2}\omega^{2}(u)\left(\bar{x} - \delta_{\gamma}\right)^{2} \nonumber\\
  & + & \gamma(u)\left(\bar{x} - \delta_{\gamma}\right) + w(u) \nonumber\\
  & + & \frac{m(u)}{2}\omega^{2}(u)\alpha \nonumber\\
  \langle\langle V^{\prime}(\bar{x} - \delta_{\gamma} + \xi) \rangle\rangle & = &
  m(u)\omega^{2}(u)\left(\bar{x} - \delta_{\gamma}\right) + \gamma(u) \nonumber\\
  \langle\langle V^{\prime\prime}(\bar{x} - \delta_{\gamma} + \xi) \rangle\rangle & = &
  m(u)\omega^{2}(u)~.
\end{eqnarray}
Comparing with (\ref{eq:SelfConsistentV_0}), we find the parameters
\begin{eqnarray}
  \omega(u; \bar{x}) & = & \omega(u) \nonumber\\
  \gamma(u; \bar{x}) & = & \gamma(u) + m(u)\omega^{2}(u)\bar{x} \nonumber\\
  w(u; \bar{x}) & = & V(u, \bar{x})~.
\label{eq:FHOParams}
\end{eqnarray}
Substituting (\ref{eq:FHOParams}) into (\ref{eq:V_0}), we have
$V_{0}(x, u; \bar{x}) = V(x, u)$ independent of $\bar{x}$.

Further, since $h(u)  = \nu^{-1/2}(u)$ satisfies the Pinney equation~(\ref{eq:Pinney}),
we can rewrite $\omega^{2}(u)$ as
\begin{equation}
  \omega^{2}(u) = \frac{1}{h(u)}\left(\ddot{h}(u) + \frac{1}{h^{3}(u)}\right)
  - \frac{\ddot{\mu}(u)}{\mu(u)}~.
\end{equation}
It follows that terms that arise from subsituting (\ref{eq:FHOParams}) into
(\ref{eq:ReducedDensityMatrix3}), such as
\begin{equation}
  \tilde{g}(u) = \frac{m(u) \omega^{2}(u)}{\mu(u)\sqrt{\dot{\nu}(u)}}~,
\end{equation}
may be written as
\begin{equation}
  \tilde{g}(u) = \mu(u)\left(\ddot{h}(u) + \frac{1}{h^{3}(u)}\right) - h(u)\ddot{\mu}(u)~.
\end{equation}
Hence, for example, integrating
\begin{equation}
  \tilde{G}_{a} = \int_{u_{a}}^{u_{b}} \tilde{g}(u) \sinh\left(\nu(u) - \nu_{a}\right)\,du
\end{equation}
by parts we find
\begin{eqnarray}
  \mu_{b}\sqrt{\dot{\nu}_{b}}\tilde{G}_{a} & = &
  \mu^{2}_{b}\dot{\nu}_{b}\cosh\left(\nu_{b} - \nu_{a}\right)
  - \mu_{b}\mu_{a}\sqrt{\dot{\nu}_{b}\dot{\nu}_{a}} \nonumber\\
  & - & \left(\frac{\ddot{\nu}_{b}}{2\dot{\nu}_{b}} + \frac{\dot{\mu}_{b}}{\mu_{b}}\right)
  \mu^{2}_{b}\sinh\left(\nu_{b} - \nu_{a}\right)~.
\end{eqnarray}
In addition,
it can be shown that,
for an arbitrary function $f(u)$,
\begin{eqnarray}
  \int_{u_{a}}^{u_{b}} \left[
    \tilde{f}(u) \tilde{G}_{a}(u) + \tilde{g}(u)\tilde{F}_{a}(u)\right]
  \sinh\left(\nu_{b} - \nu(u)\right)\,du = \nonumber\\
  -\mu_{b}\sqrt{\dot{\nu}_{b}}\tilde{F}_{a}
  - \mu_{a}\sqrt{\dot{\nu}_{a}}\tilde{F}_{b}
  + \int_{u_{a}}^{u_{b}} f(u)\sinh\left(\nu_{b} - \nu_{a}\right)\,du~, \nonumber\\
\end{eqnarray}
where
\begin{eqnarray}
  \tilde{f}(u) & = & \frac{f(u)}{\mu(u)\sqrt{\dot{\nu}(u)}} \nonumber \\
  \tilde{F}_{a}(u) & = & \int_{u_{a}}^{u} \tilde{f}(u^{\prime})
  \sinh\left(\nu(u^{\prime}) - \nu_{a}\right)\,du^{\prime} \nonumber \\ 
  \tilde{G}_{a}(u) & = & \int_{u_{a}}^{u} \tilde{g}(u^{\prime})
  \sinh\left(\nu(u^{\prime}) - \nu_{a}\right)\,du^{\prime}~. 
\end{eqnarray}

Using these results, we find that the reduced density matrix (\ref{eq:ReducedDensityMatrix3})
becomes a Gaussian integral in $\bar{x}$ of the form (\ref{eq:StandardGaussian}) with
\begin{eqnarray}
  A & = & -\frac{\sinh\left(\nu_{b} - \nu_{a}\right)}{4\hbar\Omega_{ab}}
  \left(u_{b} - u_{a}\right)^{2} \nonumber\\
  B & = & \frac{u_{b} - u_{a}}{2\hbar\Omega_{ab}}
  \left[\left(\mu_{b}\sqrt{\dot{\nu}_{b}}x_{b}\Omega_{a}
    + \mu_{a}\sqrt{\dot{\nu}_{a}}x_{a}\Omega_{b}
    \right)
    \right.\nonumber\\
    & - & \left(I_{1} + I_{2}\right)\Big]~.
\end{eqnarray}
Evaluating the integral (\ref{eq:Rho}), we find that $\rho(b, a)$ agrees with
(\ref{eq:FHODensityMatrix}), as expected.

\subsection{Limit cases}

\subsubsection{Constant $\mu$ and $\omega$}

The situation simplifies in the limit case that both $\mu(u)$ and $\omega(u)$ are constant.
Specifically, as noted above, we may set $\dot{\nu}(u) = \omega$ which implies that the
integrals in Eqs.~(\ref{eq:OmegaIIntDef}) may be computed analytically to give
\begin{eqnarray}
  \Omega_{a} & = & \frac{1}{\mu\omega^{3/2}}\left[\cosh 2f - 1\right] \nonumber\\
  \Omega_{b} & = & \Omega_{a} \nonumber\\
  \Omega_{ab} & = & \frac{1}{m\omega^{3}}\left[
    f \sinh 2f - \left(\cosh 2f - 1\right)\right] \nonumber\\
  I_{I} + I_{2} & = & \frac{\sinh 2f}{\mu\omega^{3/2}}\int_{u_{a}}^{u_{b}} \tilde{\gamma}(u)\,du
  - \frac{1}{\mu\omega^{3/2}}\left(\tilde{\Gamma}_{a} + \tilde{\Gamma}_{b}\right)~,
  \nonumber\\
\label{eq:OmegaConstMuOmega}
\end{eqnarray}
where we have defined $f \equiv \omega\left(u_{b} - u_{a}\right) / 2$.
Using the identity $\tanh f \equiv (\cosh 2f - 1) / \sinh 2f$,
we have
\begin{equation}
  \frac{\Omega_{a}}{\Omega_{ab}} = \frac{\mu\omega^{3/2}\tanh f}{f - \tanh f}
\end{equation}
and, hence,
\begin{eqnarray}
  \alpha & = & \frac{\hbar}{2m\omega}\left(\coth f - \frac{1}{f}\right) \nonumber\\
  \delta_{\gamma} & = & \frac{1}{2m\omega} \left[
    \left(\frac{\Gamma_{a} + \Gamma_{b}}{\sinh 2f}\right) \coth f
    - \frac{\hat{\gamma}}{f}\right]~,
\label{eq:AlphaDeltaConstMuOmega}
\end{eqnarray}
where we have introduced the shorthand notations
\begin{eqnarray}
  \hat{\gamma} & = & \int_{u_{a}}^{u_{b}} \gamma(u)\,du \nonumber \\
  \Gamma_{a} = \mu\sqrt{\omega}\tilde{\Gamma}_{a}, \
  \Gamma_{b} & = & \mu\sqrt{\omega}\tilde{\Gamma}_{b}, \
  \Gamma_{ab} = m\omega\tilde{\Gamma}_{ab}~.  
\end{eqnarray}
Substituting (\ref{eq:OmegaConstMuOmega}) into (\ref{eq:ReducedDensityMatrix3}) and simplifying,
we obtain
\begin{eqnarray}
  \bar{\rho}_{0}(b, a; \bar{x}) & = &
  \left(\frac{m}{2\pi\hbar\left(u_{b} - u_{a}\right)}\right)^{1/2}
  \frac{f}{\sinh f}\frac{1}{\sqrt{2\pi\alpha}} \nonumber \\
  & \times & \exp\left\{-\frac{1}{2\alpha}\left(\frac{x_{b} + x_{a}}{2} -
    \left(\bar{x} - \delta\right)\right)^{2}
    \right. \nonumber\\
    & - & \frac{m\omega\coth f}{4\hbar}\left(x_{b} - x_{a}\right)^{2}
    - \frac{1}{\hbar}\int_{u_{a}}^{u_{b}} w(u)\,du \nonumber\\
    & - & \left(x_{b} - \bar{x}\right)\frac{\Gamma_{a}}{\hbar \sinh 2f}
    - \left(x_{a} - \bar{x}\right)\frac{\Gamma_{b}}{\hbar \sinh 2f} \nonumber\\
    & + & \left.
    \frac{1}{\hbar m \omega}\left[
    \frac{\Gamma_{ab}}{\sinh 2f}
    - \frac{1}{4 f}\left(\frac{\Gamma_{a} + \Gamma_{b}}{\sinh 2f}
    - \hat{\gamma}\right)^{2}
    \right]
    \right\}~,
  \nonumber\\
\label{eq:ReducedDensityMatrixConstMuOmega}
\end{eqnarray}
with
\begin{equation}
  \delta = \frac{1}{2m\omega f}\left(
  \frac{\Gamma_{a} + \Gamma_{b}}{\sinh 2f} - \hat{\gamma}\right)~.
\end{equation}
This is the result used in Parker  {\em et al.} \cite{ParkerStedmanCapriotti2023}.

\subsubsection{Constant $\mu$, $\omega$, and $\gamma$}

In the case the function $\gamma(u)$ is also constant, the integrals in Eq.~(\ref{eq:GammaIntDef}) may also
be computed analytically to give $\tilde{\Gamma}_{a} = \tilde{\Gamma}_{b} = \gamma\Omega_{a}$ and
$\tilde{\Gamma}_{ab} = \gamma^{2}\Omega_{ab}$.  Substituting into
(\ref{eq:AlphaDeltaConstMuOmega}), we find that $\delta_{\gamma} = 0$ for all $\gamma$.

Further, substituting into (\ref{eq:ReducedDensityMatrixConstMuOmega}), we find that all terms
in $\gamma$ cancel and we are left with 
\begin{eqnarray}
  \bar{\rho}_{0}(b, a; \bar{x}) & = &
  \left(\frac{m}{2\pi\hbar\left(u_{b} - u_{a}\right)}\right)^{1/2}
  \frac{f}{\sinh f}\frac{1}{\sqrt{2\pi\alpha}} \nonumber \\
  & \times & \exp\left\{-\frac{1}{2\alpha}\left(\frac{x_{b} + x_{a}}{2} - \bar{x}\right)^{2}
    \right. \nonumber\\
    & - & \frac{m\omega\coth f}{4\hbar}\left(x_{b} - x_{a}\right)^{2} 
    - \left. \frac{1}{\hbar}\int_{u_{a}}^{u_{b}} w(u)\,du\right\}~.
  \nonumber\\
\label{eq:ReducedDensityMatrixConstMuOmegaGamma}
\end{eqnarray}
Hence, identifying $u_{a} = 0$ and $u_{b} = \beta \hbar$, we see that our extended GTFK
approximation reduces to the standard GTFK approximation \cite{Cuccoli1995}.  Note that in this
case, the second self-consistent equation in Eq.~(\ref{eq:SelfConsistent}) may be omitted.

\section{Generalized short rate models}
\label{sec:GeneralizedShortRate}

The defining SDE of the class of one-dimensional generalized short rate models is
\begin{equation}
  dX(t) = \kappa(t)\left(\theta(t) - X(t)\right)\,dt + \sigma(t)\,dW(t)~,
\label{eq:SDEShortRate}
\end{equation}
where $X(0) = x_{0}$, $\kappa(t)$ is the mean reversion speed, $\theta(t)$ is the mean reversion
level, $\sigma(t)$ is the volatility, and $W(t)$ is a standard Brownain motion.
The time-dependence of the mean reversion level is a key feature of this class of modes that is
necessary to calibrate the dynamics to the initial term structure of interest or hazard rates.

From Eqs.~(\ref{eq:PsiPathInt}) and (\ref{eq:PsiGenHamiltonian}), the generalized AD density
may be represented as the path integral (\ref{eq:PsiPathInt}) with the Hamiltonian
\begin{eqnarray}
  H(x, \dot{x}, u) & = & \frac{1}{2\sigma^{2}(u)}\left[
    \dot{x} - \kappa(u)\left(\theta(u) - x\right)\right]^{2} - \frac{\kappa(u)}{2} \nonumber\\
    & + & \lambda r(x, u)~,
\label{eq:PsiGSRHamiltonian}
\end{eqnarray}
where the precise form of $r(x, u)$ depends on the specific model.

Inserting (\ref{eq:PsiGSRHamiltonian}) into (\ref{eq:PsiPathInt}) and integrating the terms in
$\dot{x}x$ and $\dot{x}$ by parts, we obtain
\begin{eqnarray}
  \psi_{\lambda}(b, a) = e^{-W(b, a)} \rho_{\lambda}(b, a)~,
\label{eq:GSRPsi}
\end{eqnarray}
where
\begin{eqnarray}
  W(b, a) & = & \frac{1}{2}\left[m(u_{b})\kappa(u_{b})x_{b}^{2}
    - m(u_{a})\kappa(u_{a})x_{a}^{2}\right] \nonumber\\
  & - & \left[m(u_{b})\kappa(u_{b})\theta(u_{b})x_{b}
    - m(u_{a})\kappa(u_{a})\theta(u_{a})x_{a}\right] \nonumber\\
\end{eqnarray}
depends on the end points only and
\begin{equation}
  \rho_{\lambda}(b,  a) = \int_{x(a) = x_{a}}^{x(b) = x_{b}} \hspace{-0.1cm} 
  e^{-\int_{u_{a}}^{u_{b}} H_{\lambda}(x(u), \dot{x}(u), u)\,du}
  \, \,{\cal D} [x(u)]~,
\label{eq:GSRRho}
\end{equation}
where
\begin{eqnarray}
  H_{\lambda}(x, \dot{x}, u) & = & a(u)\dot{x}^{2} + \left(c(u) - \dot{b}(u)\right)x^{2}
  \nonumber\\
  & + & 2\left(e(u) - \dot{d}(u)\right)x + f(u) \nonumber\\
  & + & \lambda r(x, u) 
\label{eq:GSRHamiltonian}
\end{eqnarray}
and
\begin{eqnarray}
  a(u) = \frac{1}{2\sigma^{2}(u)} & , &
  b(u) = \frac{\kappa(u)}{2\sigma^{2}(u)} \nonumber\\
  c(u) = \frac{\kappa^{2}(u)}{2\sigma^{2}(u)} & , &
  d(u) = -\frac{\kappa(u)\theta(u)}{2\sigma^{2}(u)}   \nonumber\\
  e(u) = -\frac{\kappa^{2}(u)\theta(u)}{2\sigma^{2}(u)} & , &
  f(u) = \frac{\kappa^{2}(u)\theta^{2}(u)}{2\sigma^{2}(u)} - \frac{\kappa(u)}{2}~.\nonumber\\
\end{eqnarray}

Under the GTFK approximation, the time-$u_{a}$ value of a European derivative with expiry
$u_{b} \ge u_{a}$ may be written as
\begin{equation}
  V(u_{a}) = \int_{-\infty}^{\infty} \int_{-\infty}^{\infty} e^{-W(b, a)}
  \bar{\rho}_{0}(b, a; \bar{x}) P(x_{b})\,dx_{b}\,d\bar{x}~.
\end{equation}
Since, $W(b, a)$ is quadratic in $x_{b}$, the integral over $x_{b}$ is a Gaussian integral of
the form (\ref{eq:StandardGaussian}) and may be calculated analytically to obtain
\begin{equation}
  V(u_{a}) = \int_{-\infty}^{\infty}
  N(\bar{x}) e^{\frac{1}{4A(\bar x)} \partial^2_x}
  \left . P\left ( x \right)\right|_{x = \frac{B(\bar x)}{2A(\bar x)} }
  \,d\bar{x}~.
\label{eq:Value}
\end{equation}
where
\begin{eqnarray}
  A(\bar{x}) & = & \frac{\mu^{2}_{b}\dot{\nu}_{b}}{2}\left[
    \frac{\kappa(u_{b})}{\dot{\nu}_{b}} - \left(\frac{\ddot{\nu}_{b}}{2\dot{\nu}_{b}^{2}}
    + \frac{\dot{\mu}_{b}}{\mu_{b}\dot{\nu}_{b}}\right)
    \right.\nonumber\\
    & + & \left. \frac{1}{\sinh\left(\nu_{b} - \nu_{a}\right)}
    \left(\cosh\left(\nu_{b} - \nu_{a}\right)
    + \frac{\Omega_{a}^{2}}{2 \Omega_{ab}}\right)
    \right] \nonumber \\
  B(\bar{x}) & = & \mu_{b}\sqrt{\dot{\nu}_{b}}\left[
    \frac{\mu_{b}\kappa(u_{b})\theta(u_{b})}{\sqrt{\dot{\nu}_{b}}}
    - \left(\frac{\ddot{\nu}_{b}}{2\dot{\nu}_{b}^{2}}
    + \frac{\dot{\mu}_{b}}{\mu_{b}\dot{\nu}_{b}}\right)\bar{x}_{b}
    \right. \nonumber\\
  & + & \frac{1}{\sinh\left(\nu_{b} - \nu_{a}\right)}
    \left[\bar{x}_{b}\cosh\left(\nu_{b} - \nu_{a}\right)
      + \left(\tilde{x}^{\prime}_{a} - \tilde{\Gamma}_{a}\right)
    \right.\nonumber\\
    & + & \left. \left. \frac{\Omega_{a}}{2\Omega_{ab}}\left(
    \bar{x}_{b}\Omega_{a} - \tilde{x}_{a}^{\prime}\Omega_{b}
    + \left(I_{1} + I_{2}\right)\right)
    \right] \right] \nonumber\\
  N(\bar{x}) & = & \left(u_{b} - u_{a}\right)
  \left(\frac{\mu_{b}\mu_{a}\sqrt{\dot{\nu}_{b}\dot{\nu}_{a}}}
       {8 \pi \Omega_{ab} A(\bar{x})}\right)^{1/2}
       e^{C(\bar{x}) + \frac{B^{2}(\bar{x})}{4A(\bar{x})}}~,
\end{eqnarray}
$\bar{x}_{b} = \mu_{b}\sqrt{\dot{\nu}_{b}}\bar{x}$, and
\begin{eqnarray}
  C(\bar{x}) & = & \frac{1}{2}\left[
    \left(\frac{\ddot{\nu}_{b}}{2\dot{\nu}^{2}_{b}}
    + \frac{\dot{\mu}_{b}}{\mu_{b}\dot{\nu}_{b}}\right)
    \bar{x}_{b}^{2}
    -  \left(\frac{\ddot{\nu}_{a}}{2\dot{\nu}^{2}_{a}}
    + \frac{\dot{\mu}_{a}}{\mu_{a}\dot{\nu}_{a}}\right)
    \tilde{x}_{a}^{\prime2}
    \right] \nonumber\\
  & - & \frac{1}{2\sinh\left(\nu_{b} - \nu_{a}\right)}
  \Big[\left(\bar{x}_{b}^{2} + \tilde{x}_{a}^{\prime2}\right)
    \cosh\left(\nu_{b} - \nu_{a}\right)
    \nonumber\\
    & + & 2\bar{x}_{b}\left(\tilde{x}^{\prime}_{a}
    - \tilde{\Gamma}_{a}\right)
    + 2 \tilde{x}^{\prime}_{a} \tilde{\Gamma}_{b} - 2 \tilde{\Gamma}_{ab}
    \nonumber\\
    & + & \left. \frac{1}{2 \Omega_{ab}} \left(
    \tilde{x}^{\prime}_{a} \Omega_{b} - \bar{x}_{b} \Omega_{a} 
    -\left(I_{1} + I_{2}\right)\right)^{2}
    \right] \nonumber\\
  & + & \mu_{a}^{2}\kappa(u_{a})x_{a}\left(\frac{x_{a}}{2} - \theta(u_{a})\right)
  - \int_{u_{a}}^{u_{b}} w(u)\,du~.
\end{eqnarray}
Hence, for the generalized short rate models, the prices of zero-coupon bonds and European
options under the GTFK approximation may be obtained by evaluating a one-dimensional integral.

\subsection{The Gaussian short rate model}

The Hamiltonian of the Gaussian short rate model \citep{andersen2010interest, hw}
is (\ref{eq:PsiGSRHamiltonian}) with
\begin{equation}
  r(x, u) = x~.
\end{equation}
This is of the quadratic form (\ref{eq:GenHamiltonian}) and, hence, (\ref{eq:GSRPsi}) may be
calculated analytically either directly using the results in Sections~\ref{sec:GaussianPathInt}
and \ref{sec:FHODensityMatrix} or by applying the GTFK approximation in
Section~\ref{sec:GeneralizedGTFK}, which is exact in this case, to (\ref{eq:GSRRho}) and
evaluating the integral over the average point (\ref{eq:Rho}).  The result is
\begin{eqnarray}
  \psi_{\lambda}(b, a)  = e^{-W(b, a)}\rho(b, a)~,
\label{eq:HullWhitePsi}
\end{eqnarray}
where $\rho$ is given by (\ref{eq:FHODensityMatrix}), $h(u) = \dot{\nu}^{-1/2}(u)$ satisfies the
Pinney equation (\ref{eq:Pinney}), and we have defined
\begin{eqnarray}
  m(u) = \frac{1}{\sigma^{2}(u)} & , &
  \omega^{2}(u) = \kappa^{2}(u) + \frac{2\kappa(u)\dot{\sigma}(u)}{\sigma(u)}
  - \dot{\kappa}(u) \nonumber\\
  w(u) = f(u) & , &
  \gamma(u) = 2\left(e(u) - \dot{d}(u)\right) + \lambda~.
\end{eqnarray}

In this particular case, we may verify by differentiation that a solution of the Bernoulli
differential equation
\begin{equation}
  \dot{h}(u) + \left(\kappa(u) - \frac{\dot{\mu}(u)}{\mu(u)}\right)h(u) = \frac{1}{h(u)}
\label{eq:Bernoulli}
\end{equation}
is also a solution of (\ref{eq:Pinney}).  Rewriting (\ref{eq:Bernoulli}) in terms of
$\dot{\nu}(u)$, we obtain the relation
\begin{equation}
  \kappa(u) - \dot{\nu}(u) = \frac{\ddot{\nu}(u)}{2\dot{\nu}(u)}
  + \frac{\dot{\mu}(u)}{\mu(u)}~,
\label{eq:BernoulliRelation}
\end{equation}
and, thus, (\ref{eq:HullWhitePsi}) becomes
\begin{eqnarray}
    \psi_{\lambda}(b,  a) & = & \left(\frac{\mu_{b}\mu_{a}\sqrt{\dot{\nu}_{b}\dot{\nu}_{a}}}
        {2\pi \sinh\left(\nu_{b} - \nu_{a}\right)}\right)^{1/2} \nonumber\\
        & \times & \exp\left\{
        \frac{1}{2}\left[\left(1 - \coth\left(\nu_{b} - \nu_{a}\right)\right)
          \tilde{x}_{a}^{2} \right.\right. \nonumber\\
          & - & \left. \left(1 + \coth\left(\nu_{b} - \nu_{a}\right)\right)
          \tilde{x}_{b}^{2}\right] \nonumber\\
        & - & \left(\frac{\kappa(u_{a})\theta(u_{a})}{\sqrt{\dot{\nu}_{a}}}\mu_{a}
        + \frac{\tilde{\Gamma}_{b}}{\sinh\left(\nu_{b} - \nu_{a}\right)}\right)\tilde{x}_{a}
        \nonumber\\
        & + & \left(\frac{\kappa(u_{b})\theta(u_{b})}{\sqrt{\dot{\nu}_{b}}}\mu_{b}
        - \frac{\tilde{\Gamma}_{a} - \tilde{x}_{a}}
        {\sinh\left(\nu_{b} - \nu_{a}\right)}\right)\tilde{x}_{b}
        \nonumber\\
        & + & \left.\frac{\tilde{\Gamma}_{ab}}{\sinh\left(\nu_{b} - \nu_{a}\right)}
        - \int_{u_{a}}^{u_{b}} f(u)\,du\right\}~.
\label{eq:HullWhitePsi2}
\end{eqnarray}

The general solution of the Bernoulli differential equation is well known.
In the particular case of (\ref{eq:Bernoulli}), starting with the general solution, taking the
positive root, choosing the arbitrary lower integration bound to be $u_{a}$, and using
(\ref{eq:BernoulliRelation}) we obtain
\begin{equation}
  h(u) = \dot{\nu}^{-1/2}(u) = \frac{\mu(u)}{\mu_{a}\sqrt{\dot{\nu}_{a}}}
  e^{\nu(u) - \nu_{a}}
  e^{-\int_{u_{a}}^{u} \kappa(u^{\prime})\,du^{\prime}}~.  
\label{eq:HullWhiteh}
\end{equation}

In order to verify that (\ref{eq:HullWhitePsi2}) together with (\ref{eq:HullWhiteh}) defines a
valid Green's function for the Gaussian short rate model, we proceed to calculate the zero-coupon
bond price for the model, which is known in closed form.  From Eq.~(\ref{eq:zeroad}), 
the price of the zero-coupon bond between times $u_{a}$ and $u_{b}$
conditional on $x_{a}$ reads
\begin{equation}
  Z(u_{a}, u_{b}; x_{a}) = \int_{-\infty}^{\infty} \psi_{1}(b,  a)\,dx_{b}~.
\end{equation}
Since (\ref{eq:HullWhitePsi2}) is quadratic in $x_{b}$, this is a Gaussian integral of form
(\ref{eq:StandardGaussian}) and may be calculated analytically.  Evaluating the integral,
inserting (\ref{eq:HullWhiteh}), and simplifying using
several
integrations by parts, we eventually obtain
\begin{eqnarray}
  Z(u_{a}, u_{b}; x_{a}) & = & \exp\Big\{-\lambda x_{a} G(u_{a}, u_{b})
  \nonumber\\
  & - & \lambda \int_{u_{a}}^{u_{b}} \kappa(u)\theta(u)G(u, u_{b})\,du
  \nonumber\\
  & + & \left. \frac{\lambda^{2}}{2}
  \int_{u_{a}}^{u_{b}} \sigma^{2}(u) G^{2}(u, u_{b})\,du
  \right\}~, \nonumber\\
\end{eqnarray}
where
\begin{equation}
  G(u, u^{\prime}) \equiv \int_{u}^{u^{\prime}} e^{-\int_{u}^{v} \kappa(v^{\prime})\,dv^{\prime}}\,dv~.
\end{equation}
For $\lambda = 1$, this is equivalent to the well-known result for the Gaussian short rate
model which may be obtained by integrating (\ref{eq:SDEShortRate}) directly, see, e.g.,
Ref.~\onlinecite{andersen2010interest}.

\subsection{The Black Karasinski model}

The Hamiltonian of the Black Karasinski (BK) model \cite{bk} is (\ref{eq:PsiGSRHamiltonian})
with
\begin{equation}
  r(x, u) = e^{x}~,
\end{equation}
which implies that the short rate at any time horizon follows an intuitive lognormal
distribution.  This feature makes the BK model particularly suitable credit modelling since
it ensures that the default intensity is positive \cite{o2010modelling}.


Applying the generalized GTFK approximation in
Section~\ref{sec:GeneralizedGTFK} to the BK model we have, from (\ref{eq:GSRHamiltonian}),
\begin{equation}
  V(x, u) = \frac{m(u)}{2}\omega(u)x^{2} + \gamma(u)x + w(u) + \lambda e^{x}~,
\end{equation}
with
\begin{eqnarray}
  m(u) = \frac{1}{\sigma^{2}(u)} & , &
  \omega^{2}(u) = \kappa^{2}(u) + \frac{2\kappa(u)\dot{\sigma}(u)}{\sigma(u)}
  - \dot{\kappa}(u) \nonumber\\
  w(u) = f(u) & , &
  \gamma(u) = 2\left(e(u) - \dot{d}(u)\right)~.
\end{eqnarray}
Hence, we find the GTFK parameters
\begin{eqnarray}
  \omega^{2}(u; \bar{x}) & = & \omega^{2}(u)
  + \frac{\lambda}{m(u)} e^{\bar{x} - \delta_{\gamma} + \alpha / 2}
  \nonumber\\
  \gamma(u; \bar{x}) & = & m(u)\omega^{2}(u)\bar{x} + \gamma(u) \nonumber\\
  & + & m(u)\left(\omega^{2}(u; \bar{x}) - \omega^{2}(u)\right)\left(\delta_{\gamma} + 1\right)
  \nonumber\\
  w(u; \bar{x}) & = & V(\bar{x} - \delta_{\gamma}, u) \nonumber\\
  & + & m(u)\left(\omega^{2}(u; \bar{x}) - \omega^{2}(u)\right)\left(1 - \frac{\alpha}{2}\right)
  \nonumber\\
  & - & \frac{m(u)}{2}\omega^{2}(u; \bar{x})\delta_{\gamma}^{2}
  + \gamma(u; \bar{x})\delta_{\gamma} - \lambda e^{\bar{x} - \delta_{\gamma}}~,
  \nonumber\\
\label{eq:BKParams}
\end{eqnarray}
where $\alpha$ and $\delta_{\gamma}$ are given by (\ref{eq:AlphaDeltaGamma}).

The integral over $\bar{x}$ required to calculate the zero-coupon bond price or derivative value
(\ref{eq:Value}) may be calculated efficiently with adaptive quadrature methods.  For each
$\bar{x}$, it is convenient to solve the system (\ref{eq:Pinney}), (\ref{eq:BKParams}),
and (\ref{eq:AlphaDeltaGamma}) by iteration or a two-dimensional root search starting with
initial guesses for the constants $C$ and $D$ that are defined in Eqs.~(\ref{eq:CD}).  We find
empirically that starting the iteration (or root search) for a given $\bar{x}$ with the
solution for the closest $\bar{x}$ already found typically ensures convergence within a few
iterations.  For the first point, we start with the solution for $\lambda = 0$, which may be
calculated directly from the model parameters.

\begin{widetext}

  \begin{table}[tb]
\begin{tabular}{ccccc|cccc}
  \toprule
       & \multicolumn{4}{c|}{Typical Volatility} & \multicolumn{4}{c}{High Volatility} \\
  $T$  & GTFK   & PDE    & Abs.~Diff. & Rel.~Diff. & GTFK   & PDE    & Abs.~Diff. & Rel.~Diff.\\
  \colrule
  0.1  & 0.9940 & 0.9940 & 0.0000    & 0.00\%    & 0.9939 & 0.9939 & 0.0000    & 0.00\% \\
  0.5  & 0.9696 & 0.9696 & 0.0000    & 0.00\%    & 0.9667 & 0.9667 & 0.0000    & 0.00\% \\
  1.0  & 0.9386 & 0.9386 & 0.0000    & 0.00\%    & 0.9271 & 0.9270 & 0.0000    & 0.00\% \\
  2.0  & 0.8756 & 0.8756 & 0.0000    & 0.00\%    & 0.8361 & 0.8358 & 0.0003    & 0.04\% \\
  3.0  & 0.8132 & 0.8132 & 0.0000    & 0.00\%    & 0.7466 & 0.7459 & 0.0007    & 0.09\% \\
  5.0  & 0.6976 & 0.6975 & 0.0002    & 0.02\%    & 0.6054 & 0.6041 & 0.0013    & 0.21\% \\
  10.0 & 0.4897 & 0.4891 & 0.0006    & 0.13\%    & 0.4173 & 0.4153 & 0.0019    & 0.46\% \\
  20.0 & 0.2946 & 0.2930 & 0.0016    & 0.53\%    & 0.2747 & 0.2712 & 0.0035    & 1.29\% \\
  30.0 & 0.2100 & 0.2072 & 0.0028    & 1.35\%    & 0.2155 & 0.2094 & 0.0061    & 2.92\% \\
  \botrule
\end{tabular}
\caption{Black-Karasinski model $T$-year maturity zero-coupon bonds, $Z(0, T ; r_{0})$, obtained from the
  generalized GTFK approximation and by solving the associated PDE numerically.  The parameters
  of the BK process are: benchmark maturities
  $\{0.25, 0.5, 1.0, 2.0, 5.0, 10.0, 15.0, 20.0, 30.0\}$ years, $\delta u = 1 / 512$ year;
  mean reversion speed $\kappa(u_{l}) = 0.02$, $\kappa(u_{h}) = 0.01$; mean reversion level
  $\theta(u_{l}) = \ln 0.04$, $\theta(u_{h}) = \ln 0.06$; typical volatility
  $\sigma(u_{l}) = 0.5$, $\sigma(u_{h}) = 0.4$, high volatility $\sigma(u_{l}) = 1.0$,
  $\sigma(u_{h}) = 0.8$; and initial rate $r_{0} = 0.06$.}
\label{tbl:ZCB}
\end{table}  

\end{widetext}

In order to solve the Pinney equation (\ref{eq:Pinney}) with Runge Kutta methods, we rewrite it
as an equivalent system of coupled 1st order ODEs.  Instead of the usual prescription of
introducing  a new variable for $\dot{h}$, we choose the system of Bernoulli and Riccatti
differential equations augmented with (\ref{eq:h})
\begin{eqnarray}
  \dot{h}(u) & = & \frac{1}{h(u)} - \left(\kappa(u) - \frac{\dot{\mu}(u)}{\mu(u)}
  + g(u) \right)h(u) \nonumber\\
  \dot{g}(u) & = & g^{2}(u) + 2\left(\kappa(u) - \frac{\dot{\mu}(u)}{\mu(u)}\right)g(u)
  - \lambda e^{\bar{x} - \delta_{\gamma} + \alpha / 2} \nonumber\\
  \dot{\nu}(u) & = & \frac{1}{h^{2}(u)}~.
\label{eq:RKsystem}
\end{eqnarray}
This has the benefit of eliminating the dependence on $\ddot{\mu}(u)$ and, since
$\dot{h}(u) / h(u) = -\ddot{\nu}(u) / 2 \nu(u)$, enables us to obtain the value
of the function $\nu(u)$ and all of its required derivatives at a chosen set of points from a
single Runge Kutta problem.  We impose the boundary conditions
$h(u_{a}) = \omega^{-1/2}(u_{a})$ and $\dot{h}(u_{a}) = 0$, which implies
\begin{equation}
  g(u_{a}) = \left(\kappa^{2}(u_{a}) - \frac{\lambda}{\mu_{a}^{2}}
    e^{\bar{x} - \delta_{\gamma} + \alpha / 2} \right)^{1/2}
  - \kappa(u_{a})~.
\end{equation}

As a concrete example, we consider the general case in which all of $\mu(u)$, $\omega(u)$,
$\gamma(u)$, and $w(u)$ are time-dependent step functions with steps at $u_{i}$, $i = 1, 2,
\ldots$, as is typical of the output of a model calibration routine when calibrating to a
set of benchmark instruments~\cite{andersen2010interest}.
To simplify the problem definition, we construct the step functions
for each model parameter by specifying the values at the shortest, $u_{l}$, and longest,
$u_{h}$, benchmark maturities and obtain the values at the intermediate benchmark maturities
by linear interpolation.

In order to ensure that the required derivatives exist, we smooth the
step functions by replacing each step with a cubic.  Specifically, we replace the step at
$u_{i}$ with a cubic in the region $[u_{i}, u_{i} + \delta u]$ with parameters chosen such that
the value and first derivative are continuous at $u_{i}$ and $u_{i} + \delta u$.  Thus, we
obtain a set of intervals $\{I_{k}\}$, $k = 0, 1, \ldots$, bounded by the set of knot points
$u_{a} = u_{0} < u_{1} < u_{1} + \delta u < u_{2} < \ldots < u_{b}$.

Given the solution to the system (\ref{eq:RKsystem}) for a given $\bar{x}$, the integrals
required to obtain $C$, $D$, and the integrand in Eq.~(\ref{eq:Value}) may be computed with a
low-degree quadrature method in each interval $I_{k}$.  In practice, we obtain the required
quadrature abscissae directly from the Runge Kutta method, thus avoiding any additional
approximation due to an exogenous choice of interpolation scheme.  This is particuarly efficient
for Runge Kutta methods, such as those obtained by Bogacki and Shampine~\cite{Bogacki1996},
which support interpolation.

Table~\ref{tbl:ZCB} compares the generalized GTFK approximation results for $T$-year
maturity zero-coupon bonds in the Black Karasinski model in both typical and high volatility
market environments with those obtained by solving the associated PDE numerically.  We see
that the generalized GTFK approximation provides accurate results even in regimes of high
volatility and multi-year time horizons.

\section{Conclusion}
\label{sec:Conclusion}

We have generalized the GTFK approximation in quantum statistical mechanics to time-dependent
Hamiltonians, thus extending the scope of the method to other fields, and have demonstrated its
effectiveness for a class of generalized short rate models that are popular in mathematical
finance.

The new approximation provides remarkably accurate results for the Black-Karasinski model for
interest rates or default intensities, even for high volatilities and long time horizons, with
results that compare favorably with previously presented approximation schemes \cite{Hagan07,
  AntonovSpector2011, daniluk2016, EEBK, horvath2018analytic} and expressions that are
more compact, easier to compute, and with less severe limitations arising from, e.g., a
finite convergence radius in the time to maturity or volatility.

The accuracy and ease of computation of the GTFK method makes it a computationally efficient
alternative to fully-numerical schemes such as PDE solvers or Monte Carlo simulation. 
This is of practical utility, e.g.,  for econometric applications \cite{sahalia1999}, and in a variety of derivatives pricing applications,
such as the calculation of conditional discount factors and survival probabilities in the context of multi-factor simulations, e.g., for XVA
\cite{LeeCapriotti2015}, or the calculation of CDS quanto corrections\cite{Mercurio2019}.  Furthermore, although we have presented the GTFK approximation
in the context of short rate and default intensity models, the method could be applied to other
problems such as Asian options  \cite{devreese2010}, options on realized variance, and
stochastic volatility models.

Finally, while we have focused on an a specific mathematical finance application, the generalized GTFK method may also have applications 
in other fields where it is of interest to obtain an accurate approximation of the Fokker-Planck equation with time dependent parameters.

\appendix

\bibliography{biblio}     

\section{General one-dimensional Gaussian path integrals}
\label{sec:GaussianPathInt}

Here we state the translation of the results of Grosjean and Goovaerts \cite{GrosjeanGoovaerts1988}
to the case Euclidean path integrals.  The proof follows that for the corresponding quantum mechanical case with the exception
that the derivation of the normalization constant uses the standard Gaussian integral
\begin{equation}
  \int_{-\infty}^{\infty} e^{Ax^{2} + Bx}\,dx = \sqrt{\frac{\pi}{-A}} e^{-B^{2} / 4A},
  \ \mathrm{Re}[A] < 0~
\label{eq:StandardGaussian}
\end{equation}
rather than the Fresnel integrals.

The most general form of Hamiltonian for which (\ref{eq:DensityMatrix}) is a Gaussian path
integral is
\begin{eqnarray}
  H(x, \dot{x}, u) & \equiv & a(u)\dot{x}^{2} + 2b(u)\dot{x}{x} + c(u)x^{2} \nonumber\\
                   & + & 2d(u)\dot{x} + 2e(u)x + f(u)~,
\label{eq:GenHamiltonian}
\end{eqnarray}
where the six coefficients are continuous functions of time and $a(u)$, $b(u)$, and $d(u)$ have
continuous derivatives.  In this case, Eq.~(\ref{eq:DensityMatrix}) may be evaluated analytically
to give
\begin{equation}
  \rho(b,  a) = \frac{1}{\sqrt{\pi\hbar R(u_{b}, u_{a})}} e^{-S_{\mathrm{cl}}(b, a) / \hbar}~,
\label{eq:GenDensityMatrix}
\end{equation}
where
\begin{eqnarray}
  S_{\mathrm{cl}}(b, a) & = & \left(a(u_{b})\dot{x}_{cl}(u_{b}) + b(u_{b})x_{b}
  + 2d(u_{b})\right)x_{b}
  \nonumber\\
  & - & \left(a(u_{a})\dot{x}_{cl}(u_{a}) + b(u_{a})x_{a} + 2d(u_{a})\right)x_{a}
  \nonumber\\
  & + & \int_{u_{a}}^{u_{b}} \left[\left(e(u) - \dot{d}(u)\right)x_{\mathrm{cl}}(u)
    + f(u)\right]\,du
  \nonumber\\  
\label{eq:GenS_cl}
\end{eqnarray}
is the {\em classical action}, $x_{\mathrm{cl}}(u)$ is the {\em classical path} which satisfies
the Euler equation of motion
\begin{equation}
  \frac{\partial}{\partial u} \frac{\partial H}{\partial \dot{x}}
  - \frac{\partial H}{\partial x} = 0
\end{equation}
with $x(u_{a}) = x_{a}$, $x(u_{b}) = x_{b}$, and  $R(u, u^{\prime})$ is defined as
\begin{equation}
  R(u, u^{\prime}) = y_{2}(u)y_{1}(u^{\prime}) - y_{1}(u)y_{2}(u^{\prime})~.
\end{equation}

Further, $x_{\mathrm{cl}}(u)$ may be written as
\begin{equation}
  x_{\mathrm{cl}}(u) = C_{1}(u_{b}, u_{a})y_{1}(u) + C_{2}(u_{b}, u_{a})y_{2}(u) + z(u)~,
\end{equation}
where $y_{1}(u)$ is a non-trivial particular solution of the homogeneous equation of motion
\begin{equation}
  a(u)\ddot{x} + \dot{a}(u)\dot{x} + \left(\dot{b}(u) - c(u)\right)x = 0~,
\label{eq:GenHomogeneousEqnOfMotion}
\end{equation}
$y_{2}(u)$ is another solution of (\ref{eq:GenHomogeneousEqnOfMotion}) given by
\begin{equation}
  y_{2}(u) = y_{1}(u) \int_{\tau}^{u} \frac{du^{\prime}}{a(u^{\prime})y_{1}^{2}(u^{\prime})}~,
\label{eq:y_2}
\end{equation}
where $\tau$ is an arbitrary-chosen lower integration bound, and 
\begin{eqnarray}
  C_{1}(u_{b}, u_{a}) & = & \frac{(x_{a} - z(u_{a}))y_{2}(u_{b}) - (x_{b} - z(u_{b}))y_{2}(u_{a})}
  {R(u_{b}, u_{a})} \nonumber\\
  C_{2}(u_{b}, u_{a}) & = & \frac{(x_{b} - z(u_{b}))y_{1}(u_{a}) - (x_{a} - z(u_{a}))y_{1}(u_{b})}
  {R(u_{b}, u_{a})} \nonumber\\
  z(u) & = & \int_{\tau}^{u} \left(e(u^{\prime}) - \dot{d}(u^{\prime})\right) R(u, u^{\prime})
  \,du^{\prime}~.
\end{eqnarray}

\section{Derivation of the density matrix for the forced harmonic oscillator with
  time-dependent parameters}
  \label{sec:tdfho}

The equation of motion corresponding to the Hamiltonian in Eq.~(\ref{eq:FHOHamiltonian}) is
\begin{equation}
   m(u)\ddot{x} + \dot{m}(u)\dot{x} - m(u)\omega^{2}(u)x = \gamma(u)~,
\end{equation}
with $x(u_{a}) = x_{a}$ and $x(u_{b}) = x_{b}$.  Defining $\mu(u) \equiv \sqrt{m(u)}$ and taking
the positive root, we may write the homogeneous equation of motion as
\begin{equation}
   \ddot{x} + 2\frac{\dot{\mu}(u)}{\mu(u)}\dot{x} - \omega^{2}(u)x = 0~.
\label{eq:FHOHomogeneousEqnOfMotion}
\end{equation}

Applying the results of Section~\ref{sec:GaussianPathInt},
we obtain the classical action
\begin{widetext}
\begin{eqnarray}
  S_{\mathrm{cl}}(b, a) & = & \frac{1}{R(u_{b}, u_{a})}
  \left[\frac{m(u_{b})x_{b}^{2}}{2}\left(
    \dot{y}_{2}(u_{b})y_{1}(u_{a}) - \dot{y}_{1}(u_{b})y_{2}(u_{a})
    \right) 
    - 2x_{b}x_{a} 
    + \frac{m(u_{a})x_{a}^{2}}{2}\left(
    \dot{y}_{2}(u_{a})y_{1}(u_{b}) - \dot{y}_{1}(u_{a})y_{2}(u_{b})
    \right) \right.\nonumber\\
    & + & \left(x_{b}y_{1}(u_{a}) - x_{a}y_{1}(u_{b})\right)
    \int_{u_{a}}^{u_{b}}\gamma(u)y_{2}(u)\,du
    - \left(x_{b}y_{2}(u_{a}) - x_{a}y_{2}(u_{b})\right)
    \int_{u_{a}}^{u_{b}}\gamma(u)y_{1}(u)\,du \nonumber\\
    & + & \frac{1}{2}y_{1}(u_{b})y_{1}(u_{a}) 
    \int_{u_{a}}^{u_{b}}\gamma(u)y_{2}(u)
    \int_{u_{a}}^{u}\gamma(u^{\prime})y_{2}(u^{\prime})\,du^{\prime}du
    + \frac{1}{2}y_{2}(u_{b})y_{2}(u_{a})
    \int_{u_{a}}^{u_{b}}\gamma(u)y_{1}(u)
    \int_{u_{a}}^{u}\gamma(u^{\prime})y_{1}(u^{\prime})\,du^{\prime}du \nonumber\\
    & - & \frac{1}{2}y_{1}(u_{b})y_{2}(u_{a})
    \int_{u_{a}}^{u_{b}}\gamma(u)y_{2}(u)
    \int_{u_{a}}^{u}\gamma(u^{\prime})y_{1}(u^{\prime})\,du^{\prime}du
    - \frac{1}{2}y_{2}(u_{b})y_{1}(u_{a})
    \left.\int_{u_{a}}^{u_{b}}\gamma(u)y_{1}(u)
    \int_{u_{a}}^{u}\gamma(u^{\prime})y_{2}(u^{\prime})\,du^{\prime}du \right]\nonumber\\
    & + & \int_{u_{a}}^{u_{b}} w(u)\,du
\label{eq:FHOS_cl1}
\end{eqnarray}
\end{widetext}

In order to simplify the defining integral of $y_{2}(u)$, we choose $y_{1}(u)$ to be of the form
\begin{equation}
  y_{1}(u) = \frac{h(u)}{\mu(u)} \cosh \nu(u)~,
\label{eq:FHOy_1a}
\end{equation}
where $h(u)$ and $\nu(u)$ are to be determined.  Substituting (\ref{eq:FHOy_1a}) into
(\ref{eq:FHOHomogeneousEqnOfMotion}) leads to the system of equations
\begin{eqnarray}
  h(u)\ddot{\nu}(u) + 2\dot{h}(u)\dot{\nu}(u) & = & 0
\label{eq:h_nu_1} \\
  \ddot{h}(u)
  + \left(\dot{\nu}^{2}(u) - \omega^{2}(u) - \frac{\ddot{\mu}(u)}{\mu(u)}\right)h(u) & = & 0~.
\label{eq:h_nu_2}
\end{eqnarray}
Eq.~(\ref{eq:h_nu_1}) implies that
\begin{equation}
  \frac{d}{du}\left(h^{2}(u)\dot{\nu}(u)\right)
  = h(u)\left(h(u)\ddot{\nu}(u) + 2\dot{h}(u)\dot{\nu}(u)\right) = 0
\end{equation}
and hence that
\begin{equation}
  h^{2}(u)\dot{\nu}(u) = \mathrm{const}~.
\end{equation}
In particular, we may choose the constant of integration to be one such that $h(u)$ is given
by (\ref{eq:h}).  With this choice, (\ref{eq:h_nu_2}) becomes the Pinney
equation~(\ref{eq:Pinney}) and
\begin{equation}
  y_{1}(u) = \frac{1}{\mu(u)\sqrt{\dot{\nu}(u)}} \cosh \nu(u)~.
\label{eq:FHOy_1b}
\end{equation}
Substituting (\ref{eq:FHOy_1b}) into (\ref{eq:y_2}) and choosing $\tau = 0$ gives
\begin{equation}
  y_{2}(u) = \frac{2}{\mu(u)\sqrt{\dot{\nu}(u)}} \sinh \nu(u)~.
\label{eq:FHOy_2}
\end{equation}
Finally, substituting for $y_{1}(u)$ and $y_{2}(u)$ in Eqs.~(\ref{eq:FHOS_cl1}) and
(\ref{eq:GenDensityMatrix}) gives the results (\ref{eq:FHODensityMatrix})-(\ref{eq:FHOS_clb}).

\end{document}